\documentclass[printer]{aa}
\usepackage[english]{babel}           
\usepackage[latin1]{inputenc} 
\usepackage{graphicx,t1enc}
\usepackage[round,authoryear]{natbib}

\begin{document}
{

\title{Near-Infrared SOAR Photometric Observations of Post Common Envelope Binaries}

\author{T. Ribeiro\inst{1} \and
  R. Baptista\inst{1}
}

\institute{Departamento de F\'isica , Universidade Federal de Santa Catarina, Campus Trindade, 88040-900, Florian\'opolis, SC, Brazil\\
  \email{tiago@astro.ufsc.br}
}

\date{Received ???; accepted ???}

\abstract 
{From a number of today known Post Common Envelopes Binaries (PCEB) only a handful has yet been observed at near-infrared (NIR) wavelengths and an even smaller number has modeled NIR light curves. At shorter wavelengths one has access to the cooler and larger components of these systems and has the chance to detect emission from its faint and heavily irradiated atmospheres. }
{By modeling NIR light curves of PCEBs we intent to constrain their system parameters and study the properties of the system components.}
{Here we present simultaneous NIR $JHK_s$ light curves of two PCEBs obtained with the $4m$ SOAR telescope.}
{
KV Vel and TW Crv are long period (P$_{\rm orb} =$ 8.6h and 7.9h, respectively) PCEBs with large irradiation effects. The results of light curve fitting provided solutions with inclination $i = (47\pm5)^\circ$, mass ratio $q = 0.3\pm0.1$ and radius of the secondary $R_2/a = 0.24^{+0.05}_{-0.03}$ (where $a$ is the orbital separation) for KV Vel, and $i = (42\pm9)\degr$, $q = 0.28\pm0.04$ and $R_2/a = 0.22\pm0.01$ for TW Crv, respectively. For KV Vel, we obtain an average value for the albedo of the secondary star of $\alpha = 0.43$, consistent in the $J$, $H$ and $K_s$-bands. For TW Crv, on the other hand, we obtain values of $\alpha_{J} = (0.4\pm0.1)$ and $\alpha_{H} = (0.3\pm0.1)$ for the $J$- and $H$-bands, respectively. 

}
{}



\titlerunning{NIR SOAR Observation of PCEBs}

\authorrunning{Ribeiro and Baptista}

\maketitle

\section{Introduction}
\label{sec:int}
{\normalsize

Cataclysmic Variables (CVs) are binary systems where a low-mass dwarf (LMD) star overfill its Roche lobe and transfers matter to a more massive white dwarf (WD) at mass transfer rates typically of  \.M $\sim 10^{-10} - 10^{-8}$ M$_\odot yr^{-1}$. Since the orbital separation  (and the Roche lobe) is expected to increase when a lower mass star transfers matter to a more massive one (see \citealt{c01} for an overview), the binary must continuously loose angular momentum to sustain mass transfer. 
%
CVs are thought to be the remnants of initially wide detached binaries where the more massive star evolves to a giant and expel its outer layers through a common-envelope (CE) configuration leaving a detached LMD+WD system. A detailed description of this evolutionary picture is given by \citet{tappert_etal07}.

Recently, a series of efforts has been made to unveil the evolutionary picture of CVs i.e, its angular-momentum loss history. Although there are alternative models, \citet{k_k_r_98} concluded that the disrupted magnetic braking model (DMB) is the most plausible explanation for CV evolution. Since the secondary star is responsible for the magnetic braking mechanism, understanding its properties is essential for the development of the DMB theory. By studying secondary stars of pre-CVs and PCEBs we are able to access stars in the same regime as those of CVs  without the accretion complexity.

In that sense we have collected near-infrared (NIR) $JHK_s$ light curves of a sample of PCEBs covering a range of orbital periods from around the period gap ($P_{orb} \sim 2-3$h) up to $P_{orb} \sim 8$h, the largest orbital period suitable for full coverage within one night of (ground) observation.  This paper reports the results of the analysis of the data from two of these systems. The observations and data reduction are presented in Sect. \ref{sec:obs_red}. Sample selection criteria and specific information about the objects of this study are given in Sect. \ref{sec:selec_samp}. By modeling the light curve of these systems we are able to put constrains on their orbital system parameters and, from the colors of each component, study their properties. In Sect. \ref{sec:ana} we present the data analysis and discussion of each target separately. Conclusions and further perspectives are presented in Sect. \ref{sec:conc}.

\section{Observations and reduction}
\label{sec:obs_red}
{
All observations were performed in queue mode with the $4.1$m SOAR Telescope at Cerro Pachon, Chile, using the OSIRIS Infrared Imager and Spectrograph \citep{osiris} in imaging mode. A summary of the observations is presented in Table \ref{tab:obs}, where t$_{exp}$ is the exposure time and $ndith$ is the number of dithering positions. All runs were carried out under non-photometric conditions, with thin clouds and cirrus, resulting in variable sky transparency. Nevertheless, none of the observations had to be interrupted due to bad weather. 

Time-series of NIR photometry were performed quasi-simultaneously in $J$, $H$ and $K_s$, except in cases where the system was too faint and/or has too short orbital period that it was not possible to obtain good signal-to-noise data within the constraints given by the exposure time and the minimum time-resolution. 
The procedure of quasi-simultaneous photometry is basically to perform a set of acquisitions in one filter, change to the next filter, perform another another set and so on. This method warrants that any feature, lasting longer than  an acquisition cycle, will be present in all light curves.
\begin{table*}
 \caption{Journal of observations. Date format is yyyy-mm-dd.}              
 \label{tab:obs}      
 \centering                                      
 \begin{tabular}{c c c c c c c  } 
 \hline\hline                        
 Target & Date of & Bands & \multicolumn{3}{c}{t$_{exp}$} & $ndith$ \\ 
  name  & observation &         &$J$&$H$&$K_s$&         \\ 
 \hline                                   
 KV Vel & 2007-03-24 & $JHK_s$ & $10$s & $10$s & $20$s & $5$ \\
 TW Crv & 2007-04-24 & $JH$ & $10$s & $20$s & - & $6$ \\
 \hline                                             
 \end{tabular}
 \end{table*}

In some cases, the required total exposure time to achieve the desired signal-to-noise ratio ($S/N$) saturates the profile of the target or of a field comparison star. In these cases we have to split a single exposure into more than one, and the final image is the sum of $NC$ coadd images. For the targets discussed in this paper, the exposure time were short enough that no coadd procedure was necessary.

Data reduction was performed using IRAF\footnote{IRAF is distributed by 
the National Optical Astronomy Observatories,
which are operated by the Association of Universities for Research
in Astronomy, Inc., under cooperative agreement with the National
Science Foundation.}. All frames were first trimmed and the non-linearity of the detector was corrected by employing the 3rd order polynomial solution of \cite{osiris} for the OSIRIS instrument. The frames were then divided by the normalized flat-field. A bad pixel mask was constructed from flat-field images and used to correct for bad pixels. Since the sky contribution to NIR light is considerably high, it is standard to perform a dithering procedure, nodding the telescope into $ndith$ different positions after each image is acquired. The sky level is obtained by taking the median of the $ndith$ image. Each image is then iteratively shifted to match a reference image (usually the first image of the night) and aperture photometry is extracted for the target and all possible field stars (at least one star, aside the target, for each image field is required) using a "variable aperture" (aperture radius $= 2\times$ FWHM). The filter was changed after each set of  $ndith$ acquisitions.

Differential light curves (target star flux divided by comparison star flux) were computed in order to account for sky transparency fluctuations during the night, and flux calibrated from the {\bf 2MASS} $JHK_s$ magnitudes of the field comparison star and zero point constants \citep{2mass_cor}.
}

\section{Sample}
\label{sec:selec_samp}
{
We selected a sample of southern PCEBs without accurate NIR light curve models. We searched in the catalog of \citet{tpp_cat} for objects with confirmed orbital period of less than an observing night ($P_{orb} \leq 8$h) and for systems where the exposure time required to achieve $S/N = 50$ is less then $P_{orb}/50/N_{filters}$ (with OSIRIS+SOAR, where $N_{filters}$ is the number of filters to be used for simultaneous photometry). The later is a requirement for light curve modeling. Since we are interested in secondary stars properties "as-a-star" we do not exclude systems with subdwarf primaries. Some properties of the targets analyzed in this work are presented in Table \ref{tab:samp} and discussed below.
%
\begin{table*}
\caption{Properties of sample objects. Magnitudes were extracted from the 2MASS catalog. }              
 \label{tab:samp}      
 \centering                                      
 \begin{tabular}{c l c c c c c }          
 \hline\hline                        
  name & $P_{orb}[d]$  & $K_s$ & $J - H$ & $H - K_s$ & SP1 & SP2  \\
 \hline                                   
KV Vel	 & $0.36^{(1)}$   & $15.1^{(3)}$ & $-0.116^{(3)}$ &  $+0.644^{(3)}$ & sdOB$^{(4)}$ & M6$\pm1$V$^{(4)}$	     \\ 
TW Crv   & $0.33^{(2)}$   & $13.3^{(3)}$ & $-0.022^{(3)}$ &  $-0.016^{(3)}$ & sdO$^{(2)}$	 & M?V $^{(2)}$        	    \\
 \hline                                             
 \end{tabular}

 \tablebib{(1)~\cite{kilkenny_etat_88};
(2)~ \citet{tw_crv};
(3)~\citet{2mass_cor};
(4)~\citet{kv_discover}
}

 \end{table*}

\subsection{KV Vel}

KV Vel (or LSS 2018) is a non-eclipsing binary containing a hot subdwarf primary ($T_1 = 77~000$K) and a faint low-mass secondary star, and was the first central object of a planetary nebula discovered to be a double-lined binary \citep{kv_discover}. \citet{kv_discover} provided ephemeris, radial velocity measurements and parameter determination of the system using $UBV$ photoeletric measurements, $IUE$ spectrophotometry and high resolution spectroscopy. 
%
%

\citet{kv_discover} modeled the $UBV$ light curves of KV Vel considering a simple heating model with spherical stars and no additional flux from the secondary (besides that produced by heating). \citet{lss2018} applied the same model to UBVRI light curves of KV Vel and pointed out that the quality of the fit degraded for wavelengths longer than 5500\AA. Beyond this wavelength the secondary star contribution likely becomes significant and its distorted shape could account for the miss-fitting. In that sense, \citet{kv_vel} analysed the same data of \citet{lss2018} with an improved model that accounted for the secondary flux and distorted shape as well as for illumination and atmospheric (e.g., limb- and gravity-darkening) effects. Although this improved model was able to provide good fit to the data, a problem with the modeling illumination procedure was later reported by \citet{vi_fot_aador}. After inspection of the previous model, they found a missing factor of $\pi$ combined with a miss calculation of the surface normal vector of the emergent flux. These resulted in an underestimation, and possibly misshape, of the reflection effect, leading to unrealistic physical parameters from the fit.

As stated by \citet{lss2018} and later by \citet{kv_vel}, modeling of infrared light curves of this binary would be key to our understanding of its properties and of others of its kind, setting our motivation for this analysis.

\subsection{TW Crv}

TW Crv (or EC11575-1845) is a close binary similar to KV Vel. Both are long orbital period systems containing a hot subdwarf primary and a cool M dwarf secondary. A detailed description of this system was presented by \citet{tw_crv}, who provided the first analysis of this close binary, combining UBVRIJHK photometry and optical spectroscopy.
Their light curves were used to derive an ephemeris for the system,
\begin{equation}
\label{eq:tw_efem}
T_{mid}(E) = 2448661.6049(\pm3) + 0.32762(\pm3) \times E,
\end{equation}
by fitting a sinusoidal curve to determine the time of maximum light. A simple model of irradiated non-emitting spherical secondary, similar to that of \citet{kv_discover}, was used to model its optical light curves. As noted, this model produces poor fit to NIR light curves, since the assumption of a non-emitting secondary clearly fails at longer wavelengths.

Radial velocity curves of the system were constructed using the \ion{C}{III}/\ion{N}{III} blend at $\lambda 4650$\AA, providing velocity semi-amplitudes of $\rm K_1 = (40\pm2)$ km/s and $\rm K_2 = (119\pm4)$ km/s for the primary and for the secondary, respectively \citep{tw_crv}. As \citet{tw_crv} pointed out, the measured $K_2$ is actually the radial velocity of the irradiated face of the secondary, rather than the velocity of its center of light. They derived equations to correct that effect, as a function of the radius of the star, assuming uniform irradiation.

\citet{Exter:2005p3224} reported the analysis of high resolution ($1.1$\AA) time-resolved spectroscopy of TW Crv. They provided radial velocity measurements  for both components of the system yielding values of $\rm K_1 = (53\pm2)$ km/s and $\rm K_2 = (125\pm2)$ km/s for the primary and for the secondary, respectively. As in the case of \citet{tw_crv}, the later is also the radial velocity of the centre of light and must corrected to the center of mass of the star before being used to derive binary parameters.


For this analysis we have used new NIR light curves with higher $S/N$ and lower exposure time and an improved model to analyze the data.

\section{Data analysis}
\label{sec:ana}
{
We analyzed the data with the aid of a light curve modeling code similar to the \cite[hereafter WD]{wd_code} algorithm.  The code was developed by us to model the NIR light curves of detached and semi-detached close binaries, according to the routines outlined by \citet{KM_99}. Our code was checked against the WD code and also with the code developed by  \citet{Watson:2002p125}. It was successfully applied to derive orbital parameters of the CV IP Peg and, by extracting the light curve of the secondary, allowed the application of eclipse mapping techniques to its accretion disk \citep{rib_etal_07}. In the case of pre-CVs, the surface of the secondary is allowed to be any equipotential  within its Roche lobe for a given filling factor, which gives the gravitational potential at the surface of the star with respect to the potential of the Roche lobe ($f_{fac} = \Phi_2/\Phi_{RL1} \geq 1.0$). Thus, a filling factor of unity corresponds to a Roche lobe filling secondary star, while $f_{fac} > 1.0$ represents under-filling components.

The radiation field of the modeled component is modified to account for gravity- and limb-darkening as well as for illumination effects. The latter considers a point-like source at the position of the primary star with bolometric irradiation $I_{irr}$. We have used the gravity-darkening coefficient of \citet{grav_coef} for Roche lobe filling stars with convective envelopes ($\beta = 0.05$). Limb-darkening coefficients will be treated when discussing each target separately.

The parameters of the modeling procedure are: mass ratio ($q= M_2/M_1$), inclination ($i$), fluxes of the secondary ($F_2$) and of the primary ($F_1$), irradiated intensity ($I_{irr}$) and filling factor ($f_{fac}$). Since we have no a priori information on the orbital separation and/or primary temperature with light curve modeling, we have used $I_{irr}$ to account for the amplitude of the irradiation effect (assuming an albedo of $\alpha=1.0$) and adopted the orbital separation as our distance scale. Once we know the amount of irradiation we can discuss its origins. All light curves are modeled simultaneously. For a set of $n$ light curves, $q$, $i$ and $f_{fac}$ are common parameters while $F_2$, $F_1$ and $I_{irr}$ are particular to each data set. We proceed by minimizing the $\chi^2$ of the model with respect to the data. The procedure works by first employing a Simulated Annealing scheme to search for the region of best solution, and then an amoeba minimization routine \citep{Press:1986p3227} is employed to fine-tune the solution. Below we present and discuss the results of the modeling of the data for each target.

\subsection{KV Vel}
\label{ssec:kv}

Our NIR light curves of KV Vel (Fig. \ref{phot_kv}) are largely dominated by reflection effects, and resembles its optical light curves \citep{kv_vel}. The amplitude of the reflection effect at NIR is $\Delta H \sim 0.7$ mag against $0.55$ mag in the $V$ band. 

\begin{figure}
  \resizebox{\hsize}{!}{\includegraphics{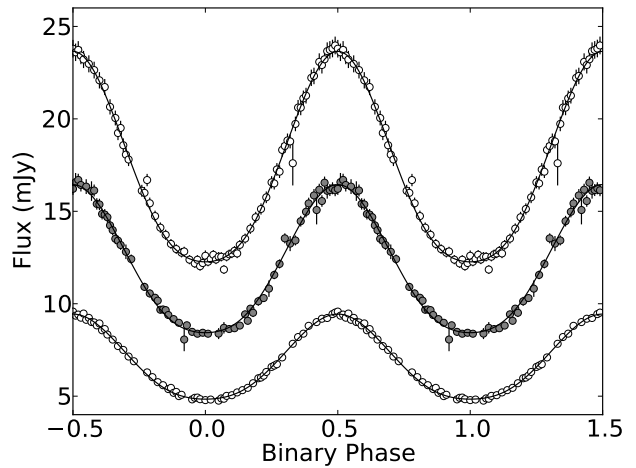}}
  \caption{Orbital light curves of KV Vel, from top to bottom in J, H and K$_s$ bands, respectively. Model light curves are overploted as solid lines. The light curves are repeated in phase for better visualization.}
  \label{phot_kv}
\end{figure}

We fitted separate sinusoids to the $\rm JHK_s$ data in order to measure the time of maximum light. The symmetric shape of the orbital hump with respect to the phase of maximum indicates that it is produced by uniform irradiation of the secondary star by a source centered at the position of the primary star. In this case, we expect that the maximum of the light curve coincides with binary phase $\phi = 0.5$, where the observer looks directly at the irradiated face of the secondary star. However, if we phase-fold the data according to the ephemeris of \cite{kilkenny_etat_88}, the maximum light is displaced from its expected phase by $\Delta \phi \sim 0.01$. The discrepancy cannot be accounted for by the uncertainty in the ephemeris of \cite{kilkenny_etat_88}.

Therefore, we combined our measured time of maximum with those of \citet[see Table \ref{tab:kv_efem}]{kv_discover, lss2018, kilkenny_etat_88} to compute a revised ephemeris for KV Vel. The best fit least-squares revised linear ephemeris is,
\begin{equation}
\label{eq:kv_ephem_new}
\rm T_{max} = HJD ~ 2\,445\,834.5174(\pm4) + 0.3571205(\pm5) \cdot E,
\end{equation}
where $\rm E$ is the cycle number. The standard deviation of the data with respect to this ephemeris is $\sigma = 9.85 \times 10^{-3}$, for a reduced chi-square of $\chi_{\nu}^2 = 1.32$. The (O-C) values with respect to this revised ephemeris are listed in Table \ref{tab:kv_efem}.

Although the revised linear ephemeris provides a nice fit to our data, it yields a poor fit to the data of \cite{kilkenny_etat_88}. These deviations prompted us to check whether a quadratic ephemeris provides a better fit to the whole data set. We applied the F-test proposed by \citet{ftest} in order to determine the statistical significance of adding an additional term to the linear ephemeris. For this case we obtain $F(1,6) = 2.8$, with a statistical significance lower than $85\%$ for the quadratic ephemeris. We conclude that presently there is no evidence of period changes in KV Vel.

\begin{table}
\caption{Times of maximum light and O-C residuals for KV Vel. Our cycle timings is the mean of individual measurements of the $JHK_s$ bands.}
\label{tab:kv_efem}
\begin{tabular}{r l c c c}
cycle & T(max)        & (O-C)\tablefootmark{a} & (O-C)\tablefootmark{b}  &  Ref. \\
          &(2400000+)&                    &                      &                             \\
 \hline 
-106   & 45796.671      & $+0.0084$ & $-0.0019$ & 1\\
   0   & 45834.52803    & $+0.0106$ & $+0.0011$ & 2 \\
   3   & 45835.59908    & $+0.0103$ & $+0.0009$ & 2\\
  829  & 46130.5735     & $+0.0032$ & $-0.0000$ & 2\\
 2834  & 46846.5850     & $-0.0118$ & $-0.0000$ & 3\\
 2845  & 46850.5130     & $-0.0122$ & $-0.0003$ & 3\\
 3018  & 46912.2940     & $-0.0130$ & $+0.0002$ & 3\\
23382  & 54184.53287    & $+0.0045$ & $-0.0092$ & 4\\
\hline                           
\end{tabular}

\tablefoottext{a}{With respect to the linear ephemeris of Eq. \ref{eq:kv_ephem_new}.}


\tablefoottext{b}{With respect to the linear ephemeris of  \citet{lss2018}.}
                    
\tablebib{(1)~\citet{kv_discover};
(2)~ \citet{lss2018};
(3)~\citet{kilkenny_etat_88};
(4) This work.}

\end{table}

We have used the previous determination of secondary star properties \citep{kv_vel} to select square root limb-darkening coefficients from \citet{limb_law}. The final set of parameters that best describe our data are shown in Table \ref{tab:kv_res}, and the corresponding light curves are shown as solid lines in Fig. \ref{phot_kv}. Error estimation was performed through a Monte-Carlo simulation.



\begin{table}
 \caption{Inferred parameters of KV Vel.}              
 \label{tab:kv_res}      
 \centering                                      
 \begin{tabular}{c c c c c c c c}          
 \hline
\hline                        
  \multicolumn{8}{c}{Fluxes of the different components.} \\
\multicolumn{4}{c}{}&         & $J$ & $H$ & $K_s$   \\
\hline
\multicolumn{4}{c}{}&$F_2(mJy)$ &$0.40\pm0.02$&$0.48\pm0.1$& $0.41\pm0.02$         \\ 
\multicolumn{4}{c}{}&$F_1(mJy)$ &$11.4\pm0.2$ &$7.6\pm0.1$& $4.2\pm0.1$\\ 
\multicolumn{4}{c}{}&$I_{irr}$ &$26.6\pm0.2$  &$15.6\pm0.1$& $10.1\pm0.1$\\ 
 \hline                                            
\multicolumn{8}{c}{Simultaneously fitted parameters:} \\
\multicolumn{8}{c}{$i = (47\pm5)^\circ$  $q =(0.3\pm0.1)$ $f_{fac}=(1.09\pm0.02)$ \tablefootmark{a}}\\
\hline
 \end{tabular}

 \tablefoottext{a}{The filling factor ($f_{fac}= \Phi_2/\Phi_{RL1} \geq 1.0$) is the relative gravitational potential at the stellar surface.}

 \end{table}

Combining the inferred inclination with the radial velocity measurements of \citet{kv_vel}, we obtain masses of $\rm M_1 = (0.7\pm0.2) M_\odot$ and $\rm M_2 = (0.3\pm0.1) M_\odot$ and an orbital separation of $a = (2.1\pm0.3) R_\odot $. Our inferred inclination is slightly lower than that of \citet{kv_vel}, leading to higher masses for both components. The inferred filling factor yields a radius\footnote{here we list $r2 = R_2/a$, the radius of the secondary in units of the orbital separation} of $r2 =  0.24^{+0.05}_{-0.03}$ or $R_2 = 0.50^{+0.07}_{-0.05} R_\odot$ for the secondary star. As noticed by \citet{kv_vel}, and underscored by our measurements, the secondary star of KV Vel is rather oversized for its mass in comparison to isolated main-sequence stars of the same mass --- likely an indication that it is out of thermal equilibrium. 

The resulting mass for the primary star of KV Vel is only marginally consistent with the canonical mass for subdwarf stars ($\rm M_{sdO} = 0.47 M_\odot$; \citealt{Han:2003p3261}). The standard scenario proposed by \citealt{Han:2003p3261} also predicts the formation of stars with masses ranging from $0.3 \rm M_\odot$ to $0.8 \rm M_\odot$. 
{\bf
In this regard, the primary star of KV Vel may correspond to the rare case of the high mass end of the predicted subdwarf distribution. Nevertheless, we point out that the radial velocity measurements of KV Vel, from \citet{kv_vel}, are potentially problematic. For instance, the lack of detailed modeling of the irradiation effect may conceivable hide important systematic errors. Therefore, further radial velocity measurements, with proper modeling of irradiation, are required to solve this issue.
}

We used the NIR colors of the low mass component of the binary to investigate its atmospheric properties and to constraint the distance to the system. The NIR colors for the secondary star of KV Vel are consistent with that of an M6V-M5V dwarf star and with those of a black body radiator with temperature T$_{\rm bb} = (3000\pm100)$K. Combining the colors and the inferred radius of the star, we infer a distance of $\rm d = (680\pm60) pc$ to the system. In addition, we have also applied the stellar evolutionary models of \cite{Baraffe:1998p637}, and  consistently obtained a distance of $\rm d = (700 \pm 100) \rm pc$ to the system. The stellar models that provided the best-fit to the NIR flux distribution of the secondary star are those of a star with a mass of ${\rm M_2} = 0.2 {\rm M_\odot}$ or ${\rm M_2} = 0.055 {\rm M_\odot}$ and ages of $ t = 10^{6.5}~ {\rm yr}$ or $t = 10^{8.5}~ {\rm yr}$, respectively. 

The resulting model properties of the stellar component are quite controversial. While and age of $10^{8.5}\, \rm yr$ is in closer agreement to that estimated by \cite{s_and_g03}, it is quite difficult to reconcile the corresponding small mass of $0.055 \rm M_\odot$ with the dynamical solution of KV Vel. For example, the discrepancy between the mass and radius of the secondary star would be even more pronounced and harder to explain. On the other hand, although a stellar component with $0.2 \rm M_\odot$ is in closer agreement with our dynamical solution, it is hard to reconcile the corresponding $10^{6.5}\, \rm yr$ age with the evolutionary stage of a PCEB. However, one might argue that, since the binary has recently evolved from a common envelope phase, the secondary star is still out of thermal equilibrium (which is also indicated by the large radius discrepancy) and, therefore, its properties resemble those of young stellar atmospheres. We conclude that the $\rm M_2 = 0.2 M_\odot$ is the most plausible solution.


\subsection{TW Crv}
\label{ssec:tw}

The light curves of TW Crv are also dominated by reflection effect, similar to those of KV Vel. Therefore, the analysis of its NIR light curve is similar to the analysis of KV Vel.

We initially phase-folded the data using the ephemeris of Eq. \ref{eq:tw_efem}. As in the case of KV Vel, the orbital minimum does not occur at phase $\phi = 0$ as expected, but rather at $\phi = -0.13$ (earlier than expected). This is not surprising given the relatively low precision of the ephemeris of \cite{tw_crv} and the long time span between their and our observations. In analogy to the previous section, we fitted a sinusoid to the light curve to measure the time of minimum light, and combined the new timing with those of \cite{tw_crv} to revise the ephemeris of TW Crv.

The best fit least-squares revised linear ephemeris for TW Crv is,
\begin{equation}
\label{eq:tw_efem_new}
\rm T_{min}(E) = 244 8661.6049(\pm3) + 0.3276074(\pm2) \cdot E.
\end{equation}

The results of the light curve fitting analysis are listed in Table \ref{tab:tw_res} and shown in Fig. \ref{fig:phot_tw}. The filling factor leads to a radius of $r_2 = 0.22\pm0.01$ for the secondary star which, using equation (5) of \citet{tw_crv}, gives a mass ratio of $q=0.30\pm0.01$, in good agreement with our light curve fitting value.

\begin{table}
 \caption{Inferred parameters of TW Crv.}              
 \label{tab:tw_res}      
 \centering                                      
 \begin{tabular}{c c c c c c c c}          
 \hline
\hline                        
  \multicolumn{8}{c}{Fluxes of the different components:} \\
\multicolumn{5}{c}{}&         & $J$ & $H$       \\
\hline
\multicolumn{5}{c}{}&$F_2(mJy)$&$2.4 \pm0.6$&$2.1\pm0.4$\\ 
\multicolumn{5}{c}{}&$F_d(mJy)$&$17.8\pm0.6$&$7.9\pm0.3$\\ 
\multicolumn{5}{c}{}&$I_{irr}$ &$11.6 \pm0.2$&$6.3\pm0.4$\\ 
 \hline                                            
\multicolumn{8}{c}{Simultaneouly fitted parameters:} \\
\multicolumn{8}{c}{$i = (41\pm9)^\circ$  $q =(0.28\pm0.04)$ $f_{fac}=(1.13\pm0.01)$ }\\
\hline
 \end{tabular}
 \end{table}

\begin{figure}
  \resizebox{\hsize}{!}{\includegraphics[angle=0]{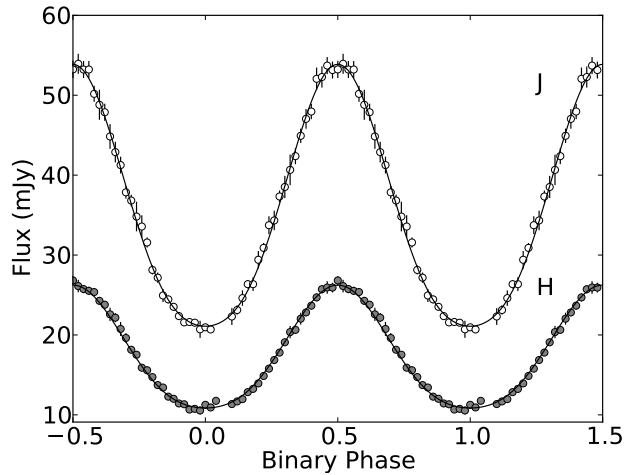}}
\caption{Light curves of TW Crv in J and H with the corresponding model fits (solid lines). The data were doubled in phase for better visualization.}
  \label{fig:phot_tw}
\end{figure}

Using eq. (4) of \citet{tw_crv} together with our values of $i$ and $r_2$ we obtain $\rm M_T =  (0.6\pm0.1) M_\odot$ for the total mass of the system, $M_1 = (0.5 \pm 0.1) M_\odot$ and $M_2 = (0.2 \pm 0.1) M_\odot$ for the stellar component masses, and an orbital separation of $a = (1.7 \pm 0.1) R_\odot$. The same calculations performed with the radial velocities  provided by \citet{Exter:2005p3224}, result in a total mass of $\rm M_T = (0.9\pm0.1) M_\odot$, $\rm M_1 = (0.66 \pm 0.05) M_\odot$ and $\rm M_2 = (0.24 \pm 0.05) M_\odot$ for the stellar component masses. In addition, we obtain an orbital separation of $a = (1.9 \pm 0.1) R_\odot$ and a radius for the secondary star of $\rm R_2 = (0.4 \pm 0.1) R_\odot$. These results are obtained by applying the same correction to K2 as in the case of the \citet{tw_crv} results.

Given the discrepancies between the results using the radial velocity values of \cite{tw_crv} and \cite{Exter:2005p3224}, it seems that the corrections derived by \citet{tw_crv} does not apply to the data of \citet{Exter:2005p3224}. For instance, a non-negligible contribution from the underlying secondary star to the narrow emission lines used by \citet{Exter:2005p3224} invalidates the procedure. A detailed modeling of the impact of irradiation to the line shape may be necessary to clarify this issue. 

The radius of the secondary star obtained from the previous analysis indicates that the secondary star is significantly oversized for its mass. In addition, the flux ratio ($\rm F_2(J)/F_2(H)$) of the secondary star is consistent with a black body of $\rm T \sim 4500K$ at a distance of $d \sim 420\, {\rm pc}$.

\section{Discussion and conclusions}
\label{sec:conc}
{
 
New NIR photometry of two long period PCEBs, KV Vel and TW Crv were presented and discussed. By measuring times of maximum light of KV Vel and TW Crv we were able to improve the system ephemerides. 

Figs. \ref{phot_kv} and \ref{fig:phot_tw}  shows our flux calibrated phase folded data together with corresponding model light curves. The model accounts for both the constant contribution of the primary\footnote{We note that the model cannot separate the constant contribution of the primary from that of any other source, like a shell or nebula, that may be contaminating the light from the system.} and of the illuminated distorted secondary, and also for inhomogeneities in the irradiation field of the secondary due to atmospheric effects. We were able to model the NIR light curves leading to well constrained system parameters (Tables \ref{tab:kv_res} and \ref{tab:tw_res}). 

From our light curve analysis it is also possible to investigate the irradiation effect on the secondary surface. In contrast to the standard procedure usually adopted for albedo modeling on these kind of binaries (p. ex., \citealt{kv_vel}), in which ``known'' physical parameters for the components are adopted and used to calculate the resulting light curve, our modeling procedure makes no {\it a-priori} assumptions on components properties. Yet, the use of previous measurements as initial set of parameters for light curve modeling is useful for accelerating the fitting procedure. Moreover, once we know the amount of irradiation with respect to the average surface intensity of the secondary ($I_s$) and, assuming an irradiating energy, we can estimate the albedo of the star. 

For this analysis, the relation of $I_s$ with the albedo can be derived from,
\begin{equation}
 \label{KM_3_2_39}
\frac{T'_{l}}{T_{l}} = \sqrt[4]{R_t}\, ,\hspace{0.5cm} \hspace{1.0cm} R_t = I_s = 1+A_t\frac{F_s}{F_t} \, ,
\end{equation}
where $T'_{l}$ is the modified effective temperature, $T_{l}$ is the local temperature at the stellar surface, $R_t$ is the local reflection factor, $A_t$ is the bolometric albedo, $F_s$ is the incident flux and $F_t$ is the undisturbed local flux \citep{KM_99}. Rearranging the terms in Eq. \ref{KM_3_2_39} we obtain,
\begin{equation}
 \label{eq:At}
A_t = \frac{I_s - 1}{F_r} \, , \hspace{1.0cm} {\rm where} \hspace{1.0cm} F_r = \frac{F_s}{F_t} \, .
\end{equation}

We adopted our inferred radius of the secondary star and the values of \citet{kv_vel} and \citet{tw_crv} for the primary stars of KV Vel and TW Crv, respectively. We proceed by considering that the primary and secondary stars in KV Vel have T$_1 = 77~000$K \citep{kv_discover} and T$_2 = 3~400$K  \citep{kv_vel}, respectively, and that they irradiate as black bodies\footnote{These assumptions are necessary in order to provide a way to measure the incident and emergent flux at the surface of the secondary.}. With these assumptions we are able to calculate the emergent ($F_t$) and incident ($F_s$) flux at the surface of the star, and complete the calculations of Eq. \ref{eq:At}. The results are listed in Table \ref{tab:albedo}.

\begin{table}
 \caption{Values for the albedo resulting from the light curve analysis.}              
 \label{tab:albedo}      
 \centering                                      
 \begin{tabular}{c c c c}          
 \hline
       &  $J$         &  $H$         &  $K_s$        \\
KV Vel & $0.43\pm0.04$  &  $0.44\pm0.04$ &  $0.41\pm0.04$  \\
TW Crv  & $0.6\pm0.1$  &  $0.5\pm0.1$ &  -   \\
\hline
\end{tabular}
\end{table}

The resulting value for the albedo is strongly dependent on the assumed temperature for both stars. Therefore, in the case of TW Crv where the temperature of the primary is not well constrained, it is difficult to estimate the albedo reliably. The results in Table \ref{tab:albedo} are obtained adopting a temperature for the primary of $T_1 = (105\pm 20) \times 10^3 \rm K$ \citep{Exter:2005p3224} and $T_2 = 4~500K$ (Sect. \ref{ssec:tw}) for the secondary.

Although the final value for the albedo still depends on the adopted temperature and irradiation field of both components (i.e. whether or not it is valid to approximate their radiation by a black body) we were able to provide a good fit to the data by adopting standard values for gravity- and limb-darkening, differently from the results of \citet{kv_vel}. Further investigation on system parameters and details of the brightness distribution of the secondary are in demand for a better understanding of the system's properties and the physics of irradiated atmospheres.

}

}

\begin{acknowledgements}
We would like to thank an anonymous referee for useful comments and suggestions, and Stella Kafka for a careful reading of an earlier version of the manuscript. This work has been done with observations from the SOAR telescope, a partnership between CNPq-Brazil, NOAO, UNC and MSU. TR acknowledges financial support from CAPES/CNPq through PhD scholarship. RB acknowledges financial support from CNPq through grant 302.442/20088. 

\end{acknowledgements}

\bibliography{../bibtex/bib}

\begin{thebibliography}{24}
\expandafter\ifx\csname natexlab\endcsname\relax\def\natexlab#1{#1}\fi

\bibitem[{Baraffe {et~al.}(1998)Baraffe, Chabrier, Allard, \&
  Hauschildt}]{Baraffe:1998p637}
Baraffe, I., Chabrier, G., Allard, F., \& Hauschildt, P.~H. 1998, A{\&}A, 337,
  403

\bibitem[{Chen {et~al.}(1995)Chen, O'Donoghue, Stobie, Kilkenny, Roberts, \&
  {van Wyk}}]{tw_crv}
Chen, A., O'Donoghue, D., Stobie, R.~S., {et~al.} 1995, MNRAS, 275, 100

\bibitem[{Claret(1992)}]{limb_law}
Claret, A. 1992, MNRAS, 335, 647

\bibitem[{Drilling(1985)}]{kv_discover}
Drilling, J. 1985, ApJ, 294, 111

\bibitem[{Exter {et~al.}(2005)Exter, Pollacco, Maxted, Napiwotzki, \&
  Bell}]{Exter:2005p3224}
Exter, K.~M., Pollacco, D.~L., Maxted, P. F.~L., Napiwotzki, R., \& Bell, S.~A.
  2005, MNRAS, 359, 315

\bibitem[{Han {et~al.}(2003)Han, Podsiadlowski, Maxted, \&
  Marsh}]{Han:2003p3261}
Han, Z., Podsiadlowski, P., Maxted, P. F.~L., \& Marsh, T.~R. 2003, Monthly
  Notice of the Royal Astronomical Society, 341, 669

\bibitem[{Hellier(2001)}]{c01}
Hellier, C. 2001, Cataclysmic Variables Stars, 1st edn. (Chichester: Springer
  and Praxis)

\bibitem[{Hilditch {et~al.}(1996)Hilditch, Harries, \& Hill}]{kv_vel}
Hilditch, R.~W., Harries, T.~J., \& Hill, G. 1996, MNRAS, 279, 1380

\bibitem[{Hilditch {et~al.}(2003)Hilditch, Kilkenny, Lynas-Gray, \&
  Hill}]{vi_fot_aador}
Hilditch, R.~W., Kilkenny, D., Lynas-Gray, A.~E., \& Hill, G. 2003, MNRAS, 344,
  644

\bibitem[{Kallrath \& Milone(1999)}]{KM_99}
Kallrath, J. \& Milone, E.~F. 1999, Eclipsing Binary Stars, 1st edn.
  (Springer-Verlag)

\bibitem[{Kilkenny {et~al.}(1988)Kilkenny, {Spencer Jones}, \&
  Marang}]{kilkenny_etat_88}
Kilkenny, D., {Spencer Jones}, J., \& Marang, F. 1988, Obs, 108, 88

\bibitem[{Kolb {et~al.}(1998)Kolb, King, \& H.}]{k_k_r_98}
Kolb, U., King, A.~R., \& H., R. 1998, MNRAS, 298, 29

\bibitem[{Kube {et~al.}(2002)Kube, Gansicke, \& Hoffmann}]{tpp_cat}
Kube, J., Gansicke, B.~T., \& Hoffmann, B. 2002, The Physics of Cataclysmic
  Variables and Related Objects (ASP Conf. Ser. 261)

\bibitem[{Landolt \& Drilling(1986)}]{lss2018}
Landolt, A. \& Drilling, J. 1986, AJ, 91, 1372

\bibitem[{Pogge {et~al.}(1999)Pogge, Martini, \& DePoy}]{osiris}
Pogge, R., Martini, P., \& DePoy, D. 1999

\bibitem[{Press {et~al.}(1986)Press, Flannery, \& Teukolsky}]{Press:1986p3227}
Press, W.~H., Flannery, B.~P., \& Teukolsky, S.~A. 1986, Cambridge: University
  Press

\bibitem[{Pringle(1975)}]{ftest}
Pringle, J. 1975, MNRAS, 170, 633

\bibitem[{Ribeiro {et~al.}(2007)Ribeiro, Baptista, Harlaftis, Dhillon, \&
  Rutten}]{rib_etal_07}
Ribeiro, T., Baptista, R., Harlaftis, E., Dhillon, V., \& Rutten, R. 2007,
  A\&A, 474, 213

\bibitem[{Sarna(1989)}]{grav_coef}
Sarna, M. 1989, A\&A, 224, 98

\bibitem[{Schreiber \& G{\"a}nsicke(2003)}]{s_and_g03}
Schreiber, M.~R. \& G{\"a}nsicke, B.~T. 2003, A{\&}A, 406, 305

\bibitem[{Skrutskie {et~al.}(2006)Skrutskie, Cutri, Stiening, Weinberg,
  Schneider, Carpenter, Beichman, Capps, Chester, Elias, Huchra, Liebert,
  Lonsdale, Monet, Price, Seitzer, Jarrett, Kirkpatrick, Gizis, Howard, Evans,
  Fowler, Fullmer, Hurt, Light, Kopan, Marsh, McCallon, Tam, S., \&
  Wheelock}]{2mass_cor}
Skrutskie, M., Cutri, R., Stiening, R., {et~al.} 2006, AJ, 131, 1163

\bibitem[{{Tappert} {et~al.}(2007){Tappert}, {G{\"a}nsicke}, {Schmidtobreick},
  {Aungwerojwit}, {Mennickent}, \& {Koester}}]{tappert_etal07}
{Tappert}, C., {G{\"a}nsicke}, B.~T., {Schmidtobreick}, L., {et~al.} 2007,
  \aap, 474, 205

\bibitem[{Watson(2002)}]{Watson:2002p125}
Watson, C. 2002, PhD Thesis

\bibitem[{Wilson \& Devinney(1971)}]{wd_code}
Wilson, R. \& Devinney, E.~J. 1971, ApJ, 166, 605

\end{thebibliography}
\bibliographystyle{aa}

\end{document}